\documentclass[conference]{IEEEtran}
\IEEEoverridecommandlockouts
\usepackage{cite}
\usepackage{amsmath,amssymb,amsfonts}
\usepackage{algorithm}
\usepackage{algorithmic}
\usepackage{graphicx}
\usepackage{textcomp}
\usepackage{xcolor}
\usepackage{caption}
\usepackage{subcaption}
\usepackage{comment}
\usepackage{multirow}
\usepackage{diagbox}
\usepackage{balance}

\usepackage{enumitem}

\usepackage{color, colortbl}
\definecolor{Gray}{gray}{0.9}
\definecolor{LightCyan}{rgb}{0.88,1,1}

\newcolumntype{a}{>{\columncolor{Gray}}c}
\newcolumntype{b}{>{\columncolor{LightCyan}}c}
\def\BibTeX{{\rm B\kern-.05em{\sc i\kern-.025em b}\kern-.08em
    T\kern-.1667em\lower.7ex\hbox{E}\kern-.125emX}}

\usepackage{pifont}
\newcommand{\cmark}{\ding{52}}%
\newcommand{\xmark}{\ding{55}}%
\begin{document}

\title{Deep Reinforcement Agent for Scheduling in HPC \thanks{The paper was accepted by IPDPS'21. If you use DRAS, please cite the paper as: Y. Fan, T. Childers, P. Rich, W. Allcock, M. Papka, and Z. Lan, ``Deep Reinforcement Agent for Scheduling in HPC", Proc. of IPDPS'21.}}

\author{\IEEEauthorblockN{Yuping Fan, Zhiling Lan}
\IEEEauthorblockA{\textit{Illinois Institute of Technology} \\
\textit{Chicago, IL}\\
yfan22@hawk.iit.edu, lan@iit.edu}
\and
\IEEEauthorblockN{Taylor Childers, Paul Rich, William Allcock}
\IEEEauthorblockA{\textit{Argonne National Laboratory} \\
\textit{Lemont, IL}\\
{\{jchilders,richp,allcock\}}@anl.gov}
\and
\IEEEauthorblockN{Michael E. Papka}
\IEEEauthorblockA{\textit{Argonne National Laboratory} \\
\textit{Northern Illinois University}\\
papka@anl.gov}
}

\maketitle

\begin{abstract}
Cluster scheduler is crucial in high-performance computing (HPC). It determines when and which user jobs should be allocated to available system resources. Existing cluster scheduling heuristics are developed by human experts based on their experience with specific HPC systems and workloads. 
However, the increasing complexity of computing systems and the highly dynamic nature of application workloads have placed tremendous burden on manually designed and tuned scheduling heuristics.
More aggressive optimization and automation are needed for cluster scheduling in HPC. 
In this work, we present an automated HPC scheduling agent named DRAS (Deep Reinforcement Agent for Scheduling) by leveraging deep reinforcement learning. DRAS is built on a novel, hierarchical neural network incorporating special HPC scheduling features such as resource reservation and backfilling. A unique training strategy is presented to enable DRAS to rapidly learn the target environment. Once being provided  a specific scheduling objective given by system manager, DRAS automatically learns to improve its policy through interaction with the scheduling environment and dynamically adjusts its policy as workload changes. 
The experiments with different production workloads demonstrate that DRAS outperforms the existing heuristic and optimization approaches by up to 45\%. 

\end{abstract}


\section{Introduction}
Cluster scheduler plays a critical role in high-performance computing (HPC). It enforces site policies through deciding when and which user jobs are allocated to system resources. Common scheduling goals include high system utilization, good user satisfaction and job prioritization. \textit{Heuristics} are the prevailing approaches in HPC cluster scheduling. For example, first come, first served (FCFS) with EASY backfilling is a well-known scheduling policy deployed on production HPC systems \cite{Feitelson02}. Bin packing is another well-known heuristic approach aiming for high utilization. Heuristics are easy to implement and fast by trading optimality for speed. 
In addition, \textit{optimization} is also extensively studied in the literature for cluster scheduling \cite{xu01,Fan3,Sun,Qiao1,Qiao2}. 
Optimization methods focus on optimizing immediate scheduling objective(s) without regard to long-term performance. 
Moreover, both heuristics and optimization approaches are static, and neither of them is capable of adapting its scheduling policy to dynamic changes in the environment.
In case of sudden variation in workloads, system administrators have to manually tune the algorithms and parameters in their policies to mitigate performance degradation.
As HPC systems become increasingly complex combined with highly diverse application workloads, such a manual process is becoming challenging, time-consuming, and error-prone. We believe that more aggressive optimization and automation, beyond the existing heuristics and optimization methods, is essential for HPC cluster scheduling.

\begin{table*}[]
\caption{Comparison of cluster scheduling methods.}
\label{scheduling_methods_comparison}
\centering
\resizebox{\linewidth}{!}{
\begin{tabular}{|l|c|c|c|c|c|}
\hline
          \backslashbox[62mm]{Features}{Methods}    & FCFS \cite{Feitelson02} & BinPacking \cite{Grandl2014} & Optimization \cite{xu01,Fan3,Sun} & Decima \cite{Mao2019} & DRAS \\ \hline
Adaption to workload changes                 &   \xmark   &    \xmark        &       \xmark       &     \cmark   &   \cmark   \\ \hline
Automatic policy tuning                      &  \xmark    &     \xmark       &     \xmark         &    \cmark    &   \cmark   \\ \hline
Long-term scheduling performance &  \xmark    &    \xmark        &     \xmark         &   \cmark     &     \cmark \\ \hline
Starvation avoidance                         &  \cmark    &      \xmark      &    \xmark          &  \xmark      &   \cmark   \\ \hline
Require training                       &  \xmark     &      \xmark      &    \xmark          &  \cmark        &   \cmark   \\ \hline
Implementation effort   &  Easy    &      Easy      &    Median        &  Hard      &   Hard   \\ \hline
Key objective   &  Fairness    &      Resource utilization      &    Customizable &  Customizable &   Customizable          \\ \hline
\end{tabular}
}
\end{table*}  

In recent years, reinforcement learning (RL) combined with deep neural networks has been successfully employed in various fields for dynamic decision making, such as self-driving cars \cite{Ahmad2017}, autonomous robots \cite{Johannink2019}, and game playing \cite{Mnih2013PlayingAW}\cite{Go2017}. Reinforcement learning refers to an area of machine learning that automatically learns to maximize cumulative reward through interaction with the environment \cite{RL2017}. Mao et al. present a RL-driven scheduling design named Decima for data processing jobs with dependent tasks \cite{Mao2019}. While Decima has shown promising results for scheduling, it is not applicable to cluster scheduling in HPC (detailed in \S \ref{Cluster Scheduling in HPC}). 

Inspired by the above RL-driven studies, we present an automated HPC scheduling agent named \textit{DRAS (Deep Reinforcement Agent for Scheduling)} tailored for HPC workloads. \textit{The goal is twofold}: (1) to improve HPC scheduling performance beyond the existing approaches, and (2) to automatically adjust scheduling policies in case of workload changes. Unlike cloud scheduling, HPC scheduling has several salient features, especially \textit{resource reservation} to prevent job starvation and \textit{backfilling} to reduce resource fragmentation. In the design of DRAS, we incorporate both features into the formulation of deep reinforcement learning and introduce \textit{a hierarchical neural network structure}, where the level-1 network selects jobs for immediate or reserved execution and the level-2 network concentrates on choosing proper backfilled jobs for more scheduling optimization. In order to optimize and automate the process, all the scheduling decisions including immediate job selection, job reservation, and backfilling are made by DRAS without human involvement. 
Moreover, we develop a three-phase training process using historical job logs. Our training strategy allows DRAS to gradually explore simple average situations to more challenging rare situations, hence leading to a fast and converged model.

We evaluate DRAS by extensive trace-based simulations with the job traces collected from two production supercomputers representing capability computing and capacity computing. The results indicate DRAS is capable of  automatically learning to improve its policy through interaction with the scheduling environment and dynamically adjusts its policy as workload changes. 
Specifically, this paper makes three major contributions: 

\begin{enumerate}[leftmargin=*]
\item We design a new scheduling agent DRAS which leverages the advance in deep reinforcement learning and incorporates the key features of HPC scheduling in the form of a hierarchical neural network model.
\item We develop a three-phase training process which allows DRAS to automatically learn the scheduling environment (i.e., the system and its workloads) and to rapidly converge to an optimal policy.
\item Our trace-based experiments demonstrate DRAS outperforms a number of scheduling methods by up to 45\%. Compared to the heuristic and optimization approaches, DRAS offers two benefits: better long-term scheduling performance and adaptation to dynamic workload changes without human intervention. 
\end{enumerate}


\section{Background and Challenges}\label{Background and Related Work}
\subsection{Cluster Scheduling in HPC}\label{Cluster Scheduling in HPC}
HPC job scheduling, also known as batch scheduling, is responsible for assigning jobs to resources (e.g, compute nodes) according to site policies and resource availability \cite{Feitelson02,Fan2,Topper}. Well-known schedulers include Slurm, Moab/TORQUE, PBS, and Cobalt \cite{SLURM,Moab,PBS,Cobalt}. Let's consider a cluster with $N$ nodes. Users submit their jobs to the system through the scheduler. When submitting a job, a user is required to provide job size $n_i$ (i.e., number of compute nodes needed for the job) and job runtime estimate $t_i$ (i.e., estimated time needed for the job). Typical HPC jobs are \textit{rigid}, meaning job size is fixed throughout its execution. Job runtime estimate is the upper bound for the job such that it will be killed by the scheduler if the actual job runtime exceeds this runtime estimate \cite{Fan1,Fan4}. At each scheduling instance, the scheduler orders the jobs in the queue according to the site policy and executes jobs from the head of the queue.

Existing HPC scheduling policies can be broadly classified into two groups: \textit{heuristics} and \textit{optimization} methods. First Come First Serve (FCFS) with EASY backfilling is the most widely used heuristics, which sorts the jobs in the wait queue according to their arrival times and executes jobs from the head of the queue. If the available resources are not sufficient for the first job in the queue, the scheduler will reserve the resources for this job. \textit{Backfilling} is often used in conjunction with reservation to enhance system utilization. It allows subsequent jobs in the wait queue to move ahead under the condition that they do not delay the existing reservations \cite{Feitelson02}. Optimization methods select a set of jobs from the queue with an objective to optimize certain scheduling metrics, such as minimizing average job wait time and maximize system utilization  \cite{xu01,Fan3,Sun,Qiao1,Qiao2}. 

Several recent studies have explored reinforcement learning for cluster scheduling. DeepRM \cite{Mao2016} is the first work demonstrating the potential of using reinforcement learning for learning customized scheduling policies from experience.
Unfortunately, DeepRM's state representation cannot handle realistic cluster workloads with continuous job arrivals. Unlike DeepRM, RLScheduler \cite{RLScheduler} attempts to develop a general reinforcement learning model that is trained with one system log and then is used on other systems with different characteristics (e.g., system size, workload patterns, etc.). While such a generic model is appealing, RLScheduler might lead to less satisfactory scheduling performance than heuristic methods.

The work closely related to ours is Decima, which explores reinforcement learning to allocate data processing jobs. Each job consists of dependent tasks and is represented as directed acyclic graphs (DAGs). Decima integrates a graph neural network to extract job DAGs and cluster status as embedding vectors. It then feeds the embedding vectors to a policy gradient network for decision making. The decision consists of two parts: to select tasks for immediate execution and to determine task parallelism. Unfortunately, Decima is not applicable to HPC scheduling. First, Decima assumes all jobs can be decomposed into malleable tasks, whereas HPC is dominated by rigid jobs that cannot be decomposed. Second, Decima can cause serious job starvation due to the lack of resource reservation support (Figure \ref{job_size_vs_wait_time}). In short, Table \ref{scheduling_methods_comparison} summarized and compared existing cluster scheduling methods, along with their features. 

\subsection{Overview of Reinforcement Learning}\label{Overview of Reinforcement Learning}

Reinforcement learning (RL) is a type of machine learning technique that studies how agents situated in stochastic environments can learn optimal policies through interaction with their environment \cite{Ipek2008}.  
The agent's environment is described by an abstraction called Markov Decision Process (MDP) with four basic components: state space $S$, action space $A$, reward $R$, and state transition probability $P$. In Markov decision processes, a learning agent interacts with a dynamic environment in discrete timesteps. At each time step $t$, the agent observes the state $s_t \in S$ and takes an action $a_t \in A(s_t)$. Upon taking the action, the environment transits to a new state $s_{t+1}$ with the transition probability $P(s_{t+1}|s_t,a_t)$ and provides a reward $r_t$ to the agent as feedback of the action. The process continues until the agent reaches a terminal state. The goal of the agent is to find a policy $\pi(s)$, mapping a state to an action (deterministic) or a probability distribution over actions (stochastic), which maximizes \textit{the long-term (discounted) cumulative reward $\sum_{t'=t}^{T} \gamma^{t'} r_{t'}$.} A discount factor $\gamma$ is between 0 and 1. The smaller of $\gamma$, the less importance of future rewards. 

In practice, the state and action space is often too large to be stored in a lookup table. It is common to use function approximators with a manageable number of adjustable parameters $\theta$, to represent the components of agents. Using a deep neural network with reinforcement learning is often called \textit{deep reinforcement learning} \cite{DL2015}. The highly representational power of deep neural networks enables reinforcement learning to solve complex decision-making problems, such as playing Atari and Go games \cite{Mnih2013PlayingAW,Go2017}.

\textit{Policy gradient} and \textit{Q-learning} are the most popular RL algorithms \cite{Sutton1999}\cite{RL2017}. 
\textit{Policy gradient} methods directly parameterize the policy $\pi_\theta(s)$ and optimize the parameters $\theta$ in the neural network by gradient descent. In Q-learning algorithms, an agent chooses an action at a given state that maximizes Q-value, i.e., the cumulative reward over all successive steps. Q-table is a lookup table containing Q-value for all the state-action pairs. To address an overwhelming number of state-action pairs, neural networks are often used to approximate Q-table and the methods are generally called \textit{deep Q-learning (DQL)}. DQL learns by approximating the optimal action-value function $ Q^*_\theta(s, a)$. Policy gradient methods are generally believed to be applicable for a wider range of problems and converge faster, but tend to converge to a local optimal. On the other hand, Q-learning methods are more difficult to converge, but once they converge, they tend to have more stable performance than policy gradient methods \cite{Mnih2016}.

\subsection{Technical Challenges}

Designing deep reinforcement learning driven cluster scheduling for HPC is challenging. Several key obstacles as listed below. 

\textbf{Avoidance of job starvation.} HPC jobs have drastically different characteristics: user jobs may range from a single-node job to a whole-system job, and job runtimes may vary from seconds to hours or even days. This feature presents a unique challenge to HPC systems: jobs, especially large-sized jobs, tend to be starved, if small-sized jobs keep arriving and skip over large jobs due to insufficient available resources. Simply applying existing RL-based scheduling methods can lead to severe job starvation. We have tested a state-of-the-art policy gradient method with a real workload trace. Our results show that large jobs, e.g., 4k-node jobs, were held in the queue for 170 days. Typically, large jobs have high priority at HPC sites, especially capability computing facilities. The long wait times discourage users from submitting large jobs.

\textbf{Incorporation of backfilling.} Backfilling is a key strategy to reduce resource fragmentation in HPC. Currently, the well-known EASY backfilling strategy uses the simple first-fit method to select jobs for backfilling, i.e., choosing the first job which can fit in the backfill hole. We argue that similar to the selection of jobs for scheduling, the selection of jobs for backfilling has many possible options, hence having the potential for more aggressive optimization. 

\textbf{Scalable state and action representation.} To transform a scheduling problem to a reinforcement learning problem, we must first capture the dynamic environment, e.g., status of thousands of nodes and hundreds of waiting jobs, to a state vector as an input to the neural network. Additionally, it is vitally important to map the extremely large action space to an output of the neural network in a manageable size. The action space grows exponentially with the number of jobs in the queue. Working directly with large action space can be computationally demanding. 

\textbf{Effective agent training.} An RL agent learns to improve its policy by experiencing diverse situations. An effective training should be capable of efficiently and rapidly building a converged model based on sample data in order to make decisions without being explicitly programmed to do so. It is also challenging to select training data to reliably cover as much of the state space as possible and generalize to new or unseen situations.

\begin{figure*}[!h]
\centerline{\includegraphics[width=0.9\linewidth]{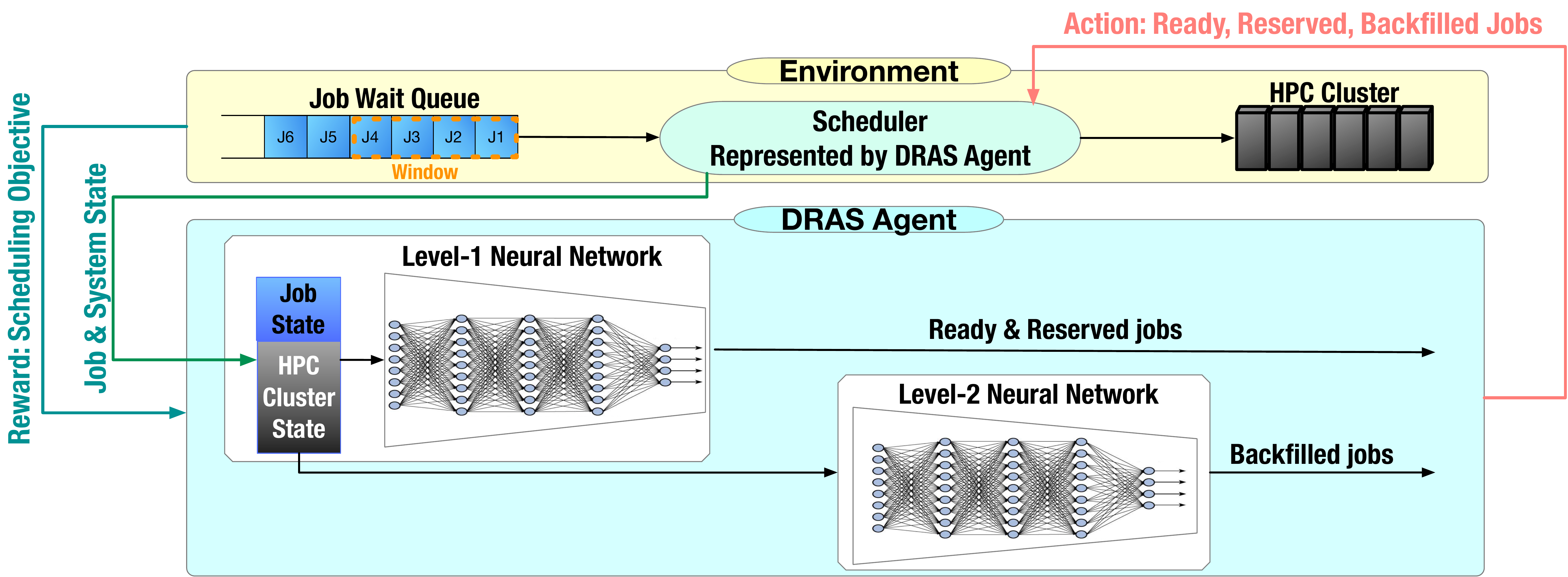}}
\caption{DRAS overview. The agent (at the bottom) represents the scheduler; the environment (at the top) comprises the rest of the system, including job wait queue and HPC cluster. The DRAS agent first observes the environment state, including job state and system state, and encodes the state into a vector. The agent's neural network takes the vector as input and outputs a scheduling action. The environment executes the action and provides a reward indicating the quality of the action. The agent uses reward to improve its policy automatically.}
\label{system_overview}
\end{figure*}

\section{Design of DRAS}\label{Methodology}

Now we present DRAS, a new scheduling method tailored for HPC workload and is empowered by deep reinforcement learning. DRAS, illustrated in Figure \ref{system_overview}, represents the scheduler as an agent to make decisions on \textit{when and which} jobs should be allocated to computer nodes with the objective to optimize scheduling performance. At a given scheduling instance $t$, the agent first encodes the job queue and system state into a vector $s_t$, and passes the vector to the neural network (\S \ref{State and Action Representation}). Next, DRAS uses a hierarchical neural network for decision making (\S \ref{Starvation and Backfill Management}). The agent takes an action by selecting jobs from the wait queue according to the output of the neural network and then receives a reward signal from the environment. The goal of DRAS is to choose actions (i.e., to select jobs) over time so as to maximize the cumulative reward.
DRAS trains its neural network through simulation with massive datasets composed of both real and synthetic workload traces (\S \ref{Training Strategy}). Once the model is converged, we deploy the DRAS agents into operation. The DRAS agents automatically adjust their neural network parameters during operation to handle workload changes.

\subsection{State, Reward and Action Representation}\label{State and Action Representation}
The DRAS agent receives three observations from the environment: (1) job wait queue, (2) cluster node status, and (3) reward, a scalar indicating the quality of the action.

\textbf{State.} We encode each waiting job as a vector of $[2,2]$, containing four pieces of information, including job size, job estimated runtime, priority (1 means high priority; 0 means low priority), and job queued time (time elapsed since submission). We encode each node as a vector of $[1,2]$ with two pieces of information. The first cell is a binary representing node availability (1 means available; 0 means not available). If the node is occupied, we use the user-supplied runtime estimate and job start time to calculate the node estimated available time. The second cell represents the time difference between the node estimated available time and the current time. If the node is available, we set the second cell to zero. We concatenate job information and node information into a fixed-size vector as the input to the neural network.

\textbf{Reward.}
Reward functions reflect scheduling objectives. It is hard to offer a one-size-fits-all reward function due to diverse site objectives. HPC systems can be broadly classified as capability computing or capacity computing. Capability computing facilities are commonly interested in prioritizing capability jobs (i.e., large jobs) \cite{Fan2} and optimizing resource utilization. An example reward of capability computing could be as follows: 
\begin{equation}
w_1 \times \frac{\overline{t_i}}{t_{max}}  
+ w_2 \times \frac{\overline{n_i}}{N} 
+ w_3 \times \frac{N_{used}}{N} \label{theta_reward_equation}
\end{equation}
where 
$\overline{t_i}$ denotes the average wait time of selected jobs; $t_{max}$ is the maximum wait time of jobs in the queue. Similarly, $\overline{n_i}$ is the average job size of the selected jobs; $N$ is the total number of nodes in the system; $N_{used}$ is the number of occupied nodes.
In other words, this reward function intends to balance three factors: to prevent job starvation, to promote capability jobs, and to improve system utilization. The weights can be tuned by system administrators based on the site priority. For example, the higher $w_1$ value could meet a more stringent requirement on job starvation.

Capacity computing facilities typically focus on fast turnaround time and short wait time \cite{CoriPolicy}.
For capacity computing facilities, we may define the reward function as:
\begin{equation}
\frac{\sum_{j \in J}  -1/t_j}{c}
\label{cori_reward_equation}
\end{equation}
where $J$ is the set of jobs in the queue and $c$ is the number of waiting jobs at the current timestep. This reward function aims to minimize the average job wait time.

\textbf{Action.} DRAS processes the input vector and outputs a vector as the scheduling action. The output vector specifies which jobs are selected for job execution (i.e., immediate execution, reserved execution, and backfilled execution). Intuitively, at each scheduling instance, the scheduler selects multiple jobs simultaneously. This leads to an explosive number of actions and is infeasible to be trained efficiently. Instead, DRAS decomposes one scheduling decision (i.e., selects several jobs in one shot) into a series of job selections, i.e., selecting one job at each time. 

\subsection{Two-level Neural Network}\label{Starvation and Backfill Management}
A key challenge when applying deep reinforcement learning to HPC cluster scheduling is to prevent job starvation. State-of-the-art RL methods focus on scheduling jobs for immediate execution and lack reservation strategy, hence leading to job starvation. To overcome this obstacle, we build \textit{a hierarchical neural network structure}, in which the level-1 network is to select jobs for immediate or reserved execution and the level-2 network is to identify jobs for backfilling. 

More specifically, at a given scheduling instance, the scheduler first enforces a window at the front of the job wait queue. The window alleviates job starvation problems by providing higher priorities to older jobs. The level-1 network selects a job from the window. If the number of available nodes is more than or equal to the job size, the agent marks the job as \textit{ready job} and sends it for immediate execution on the system. This process repeats until the job selected from the window has a size greater than the number of available nodes. The agent marks the job as \textit{reserved job} and reserves a set of nodes for its execution on the system at the earliest available time. At this point, the agent moves to the level-2 network. Unlike the first-fit strategy used in the traditional backfilling method, we use the neural network to make backfilling decisions so as to minimize resource waste. Toward this end, we fill the window with job candidates, i.e., the jobs that can be fit into the holes in the system before the reserved time. The agent selects one job at a time for system to backfill. The process at the level-2 network repeats until no more job candidates for backfilling. 

In a nutshell, the decision making of DRAS is to select jobs and execute them in three modes: 
\begin{enumerate}[leftmargin=*]
\item \textbf{ready job}: the jobs are selected to run immediately. 
\item \textbf{reserved job}: the jobs are selected to start at the earliest reserved time. 
\item \textbf{backfilled job}: the jobs are selected to fill the holes before the reserved time.
\end{enumerate}

The same neural network is used for both level-1 and level-2 networks. The entire 2-level neural network is trained jointly using deep reinforcement learning to optimize scheduling performance.  Each network consists of \textit{five layers}: input layer, convolution layer, two fully-connected layers, and output layer. The input layer is connected to a convolution layer with a $1 \times 2$ filter to extract job or node status information in each row. The convolution layer is connected to two fully-connected layers activated by leaky rectifier \cite{Goodfellow-et-al-2016}. The second fully-connected layer is connected to the output layer. We denote all of the parameters in the neural network jointly as $\theta$.

In this study, we develop two DRAS agents: \textit{DRAS-PG} and \textit{DRAS-DQL}. PG denotes policy gradient, and DQL denotes deep Q-learning. The selection of PG and DQL is for us to systematically evaluate these popular reinforcement learning methods under a unified environment.

\textbf{DRAS-PG} uses the neural network to parameterize scheduling policy as $\pi_\theta(s_k, a_k)$ (i.e., the probability of taking action $a_k$ in state $s_k$). The input of DRAS-PG is a 2D vector of $[2 \times W + N, 2]$, where $W$ is the window size and $N$ is the total number of nodes in the system. The output of the neural network contains $W$ neurons, each denoting the probability of selecting a job out of the $W$ jobs. A scheduling action is stochastically drawn from the $W$ jobs following their probability distributions.
We employ the softmax \cite{Goodfellow-et-al-2016} as the activation function to ensure the sum of output values equals to 1.0. If the number of wait jobs is less than the window size $W$, we mask the invalid actions in the output by rescaling all valid actions. 
In terms of learning, DRAS-PG method updates the neural network parameters $\theta$ by:
\begin{equation}
 \theta \leftarrow \theta + \alpha \sum_{k=1}^{K}  \bigtriangledown_{\theta} log \pi_{\theta}(s_k,a_k) (\sum_{k'=k}^{K} r_{k'} - b_k) \label{ThetaReward}
\end{equation}
Here, $K$ denotes the total number of actions taken in the parameter update, $\alpha$ is the learning rate of using Adam optimizer \cite{Adam}, and $b_k$ is the baseline used to reduce the variance of policy gradient. 
We set $b_k$  to the cumulative reward from step $k$ onwards averaging over all past parameter updates. 

\textbf{DRAS-DQL} uses the neural network to approximate Q-value as $Q_\theta(s_k,a_k)$ (i.e., the expected cumulative reward of taking action $a_k$ in state $s_k$). DRAS-DQL network processes one job at a time and produces the expected Q-value for this job. We use the same network to approximate Q-value for all the jobs in the window $W$. The input of DRAS-DQL neural network is a 2D vector of $[2 + N, 2]$, containing one job information and $N$ nodes information. The output is a single neuron corresponding to the expected Q-value of the job. After processing all the jobs in the window, normally, the agent selects the job with the highest Q-value. 

In order to explore various actions, the agent randomly chooses a job instead of the job with the highest Q-value with probability $\epsilon$. In practice, $\epsilon$ is very high at the beginning of the training to ensure that the agent explores various state-action pairs and it decays over time as the agent becomes more experienced. In our study, we set $\epsilon = 1.0$ at the beginning of the training and it decays at the rate of $\alpha = 0.995$.
In training, the parameters $\theta$ in  DRAS-DQL network is updated by:
\begin{equation}
 \theta \leftarrow \theta - \alpha \sum_{k=1}^{K}  \bigtriangledown_{\theta}Q(s_k,a_k) (\underbrace{r_k+\max\limits_a Q(s_{k+1}, a)}_{\text{new value}} -\underbrace{Q(s_k,a_k)}_{\text{old value}})    \label{CoriReward}
\end{equation}
Here, the old value $Q(s_k,a_k)$ is the expected Q-value of taking action $a_k$ at state $s_k$. After taking action $a_k$, we can compute the more accurate expected Q-value (i.e., the new value) by adding the immediate reward $r_k$ and the expected cumulative future reward. DQL networks learn through minimizing the loss between the new value and the old value.

\subsection{Training Strategy}\label{Training Strategy}
At the beginning of the training, we initialize DRAS's neural network parameters $\theta$ to random numbers. We train the neural network in episodes and the network parameters $\theta$ are updated with episodic training until convergence. For each episode, the environment is first set to its initial state (i.e., all nodes are idle and no jobs run on the system). We train DRAS via trace-based simulation, in which job events occur at a specific instant in time according to the job traces. DRAS observes the scheduling state, makes scheduling decisions according to its neural network, and collects scheduling reward. For every ten scheduling instances, DRAS updates its parameters $\theta$ based on the collected observations and then clears the memory for the next update. An episode terminates when all jobs in the jobset have been scheduled. We monitor the progress of the training by taking a snapshot of the model after each episode. The next episode uses a new jobset to refine the previous model.

The jobsets used in training determine the convergence and quality of the DRAS model. To learn a converged model, we follow the principle of gradual improvement: \textit{DRAS starts with simple average cases and gradually improves its capability with unseen rare cases.} Specifically, we train DRAS by using a three-phase training process and three types of jobsets are used to train DRAS in order: (1) a set of sampled jobs from real job traces, (2) a period of real job traces, and (3) a set of synthetic jobs generated according to job patterns on the target system.
The sampled jobsets have controlled job arrival rates providing the easiest learning environment. Once DRAS can make good scheduling decisions under the controlled environment, training on the real job traces with various job arrival patterns allows DRAS to learn more challenging situations. The final phase is to train DRAS with synthetic jobsets, which enables DRAS to experience a variety of potential states that might not be seen in the first two types of jobsets. We will show this three-phase training process leads to a fast convergence (\S \ref{DRAS Agent Training}).

DRAS is implemented in Tensorflow \cite{TensorflowURL} and available as open-source on GitHub \cite{DRASGithub}.

\section{Experimental Setup}\label{Evaluation Configuration}
\subsection{Comparison Methods}\label{comparison methods}
We compare the following scheduling methods:
\begin{itemize}[leftmargin=*]
	\item \textbf{FCFS} represents FCFS with EASY backfilling, which is the default scheduling policy deployed on many production supercomputers \cite{SLURM}. FCFS prioritizes jobs based on their arrival times and EASY backfilling is used to reduce resource fragmentation \cite{Feitelson02}.
	\item \textbf{BinPacking} is widely used heuristic method for scheduling in datacenters \cite{Grandl2014}. It iteratively allocates the largest runnable jobs (i.e., job size is less than or equal to the number of available nodes in the system) until the system cannot accommodate any further jobs.
	\item \textbf{Random} randomly selects runnable jobs from the queue to execute until no more jobs in the queue can fit into the system. Since DRAS performs similar to Random at the beginning of training by randomly explores action space, if DRAS's performance is better than Random, it demonstrates that DRAS gradually learns to improve its scheduling action.
	\item \textbf{Optimization} denotes a suite of scheduling methods that formulate cluster scheduling as an optimization problem \cite{Fan3}, e.g., to minimize average job wait time. In our experiments, the optimization problem is formulated as a 0-1 knapsack problem which is solved using dynamic programming. For a fair comparison, we use the same scheduling objectives (i.e., Equation (\ref{ThetaReward}) and (\ref{CoriReward})) for Optimization and for DRAS.
	\item \textbf{Decima-PG} denotes a modified version of Decima \cite{Mao2019}. As mentioned earlier, Decima is not designed for scheduling HPC jobs. Hence, we use a modified version of Decima by skipping graph neural network and adopting our state representation presented in \S \ref{State and Action Representation}. Note that Decima-PG is a RL agent without hierarchical network structure. Hence it acts as the baseline to demonstrate the benefits of the hierarchical design of DRAS.
	\item \textbf{DRAS-PG} and \textbf{DRAS-DQL} denote our DRAS agents. 
\end{itemize}

\subsection{Trace-based Simulation}\label{Simulation Platform}
We compare these scheduling policies through trace-based simulation. Specifically, a trace-based, event-driven scheduling simulator called CQSim is used in our experiments \cite{xu01}\cite{Li1} \cite{CQSimGithub}. CQSim contains a queue manager and a scheduler that can plug in different scheduling policies. It emulates the actual scheduling environment. A real system takes jobs from user submission, while CQSim takes jobs by reading the job arrival information in the trace. Rather than executing jobs on system, CQSim simulates the execution by advancing the simulation clock according to the job runtime information in the trace. 

\begin{table}[htp]
\centering
\caption{Theta and Cori workloads.}
\label{Cori_Theta_comp}
\resizebox{\linewidth}{!}{
\begin{tabular}{|l|l|l|}

\hline
\rowcolor{LightCyan}
                              &Theta                                                                         &  Cori            \\ \hline
\rowcolor{Gray}
Location                      &ALCF                                                                                            & NERSC                  \\ \hline
\rowcolor{LightCyan}
Scheduler             & Cobalt                                                                                            & Slurm                   \\ \hline
\rowcolor{Gray}
System Types                     & Capability computing                                                                                      & Capacity computing           \\ \hline
\rowcolor{LightCyan}
Compute Nodes                          & \begin{tabular}[c]{@{}l@{}}4,392\\ (4,392 KNL)\end{tabular} & \begin{tabular}[c]{@{}l@{}}12,076\\ (2,388 Haswell; 9,688 KNL)\end{tabular}               \\ \hline
\rowcolor{Gray}
Trace Period                  & Jan. 2018 - Dec. 2019                                                                          &Apr. 2018 - Jul. 2018   \\ \hline
\rowcolor{LightCyan}
Number of Jobs &121,837                                                                                        & 2,607,054                 \\ \hline
\rowcolor{Gray}
Max Job Length &1 day                                                                                     & 7 days                  \\ \hline
\end{tabular}
}
\end{table}

\begin{figure}[!htbp]
\begin{subfigure}{0.24\textwidth}
  \centering
  \includegraphics[width=\linewidth]{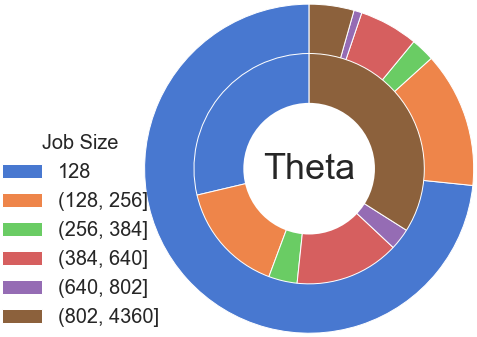}  
  \label{theta_node_distribution_pie_chart}
\end{subfigure}
\begin{subfigure}{0.24\textwidth}
  \centering
  \includegraphics[width=\linewidth]{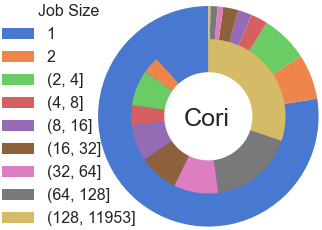}  
  \label{cori_node_distribution_pie_chart}
\end{subfigure}
\caption{Job characterization of Theta at ALCF and Cori at NERSC. The outer circle shows the number of jobs in each job size category. The inner circle presents the total core hours consumed by each job size category.}
\label{node_distribution_pie_chart}
\end{figure}

\subsection{Workload Traces}\label{simulation systems}

In our study, two real workload traces are used. Table \ref{Cori_Theta_comp} summarizes the two traces collected from production systems, and Figure \ref{node_distribution_pie_chart} gives an overview of job size distributions on these supercomputers. 
We select these traces as they represent different workload profiles: (1) capability computing  focusing on solving large-sized problems, (2) capacity computing solving a mix of small-sized and large-sized problems.
The first workload is a two-year job log from Theta \cite{Theta}, the production HPC system located at ALCF. Theta is a capability computing system. The smallest job allowed on Theta is 128-node \cite{ThetaPolicy}. Only 2.25\% of jobs have dependency. For jobs with dependency, the scheduler hides them from scheduling until all their parents have been executed. On Theta, there are 32 nodes dedicated to run debugging jobs and the rest of 4,360 nodes are dedicated to user jobs. In our experiments, we set the system size to be 4,360 and filter out all debugging jobs in the trace. \textit{We use the first 2-month data for training, the next month data for validating model convergence, and the rest 21-month data for testing.}

The second trace is a four-month job log from Cori \cite{Cori}. Cori is a capacity computing system deployed at NERSC. A majority of its jobs consume one or several nodes (Figure \ref{node_distribution_pie_chart}). 
The longest job executed for seven days. \textit{We use the first 2-week data for training, the next 1-week data for validating model convergence, and the last 15-week data for testing.}


\subsection{DRAS Training}\label{DRAS Agent Training}
The details of RL architectures for these systems are listed in Table \ref{DRAS_nn_sizes}. 
Take the neural network of \textit{DRAS-PG} on Theta as an example. The input of the neural network is a vector of $[4460, 2]$. We use a convolutional layer with 4460 neurons and two fully-connected layers with 4000 and 1000 neurons respectively. The output layer contains 50 neurons representing jobs in the window. In total, the neural network has 21,890,053 trainable parameters.

\begin{table}[!h]
\caption{DRAS network configurations for Theta and Cori.}
\label{DRAS_nn_sizes}
\resizebox{\linewidth}{!}{
\begin{tabular}{|l|l|l|l|l|}
\hline
\rowcolor{LightCyan}
\multirow{2}{*}{}       & \multicolumn{2}{l|}{Theta}       & \multicolumn{2}{l|}{Cori}          \\ \cline{2-5} 
\rowcolor{LightCyan}                        & DRAS-PG          & DRAS-DQL    & DRAS-PG           & DRAS-DQL     \\ \hline
\rowcolor{Gray}
Input                   & $[4460, 2]$ & $[4362, 2]$ & $[12176, 2]$ & $[12078, 2]$ \\ \hline
\rowcolor{LightCyan}
Convolutional Layer     & $4460$   & $4368$   & $12176$   & $12078$   \\ \hline
\rowcolor{Gray}
Fully Connected Layer 1 & \multicolumn{2}{l|}{4000}        & \multicolumn{2}{l|}{10000}         \\ \hline
\rowcolor{LightCyan}
Fully Connected Layer 2 & \multicolumn{2}{l|}{1000}        & \multicolumn{2}{l|}{4000}          \\ \hline
\rowcolor{Gray}
Output                  & 50                & 1            & 50                 & 1             \\ \hline
\rowcolor{LightCyan}
Trainable Parameters               & 21,890,053                & 21,449,004            & 161,960,053                 & 161,764,004             \\ \hline
\end{tabular}
}
\end{table}

For the capability computing facility Theta, we define its reward as Equation (\ref{theta_reward_equation}). We set the weights $w_1 = w_2 = w_3= 1/3$. For the capacity computing facility Cori, we set the reward as Equation (\ref{cori_reward_equation}). 
The learning rate $\alpha$ is set to 0.001.

\begin{figure}[!htbp]
\begin{subfigure}{0.24\textwidth}
  \centering
  \includegraphics[width=\linewidth]{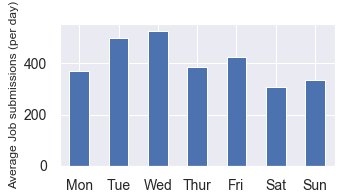}  
  \caption{Weekly job submission pattern.}
  \label{job_sub_per_count_days}
\end{subfigure}
\begin{subfigure}{0.24\textwidth}
  \centering
  \includegraphics[width=\linewidth]{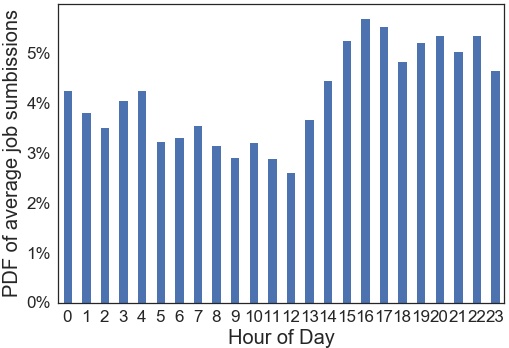}  
  \caption{Daily job submission pattern.}
  \label{job_sub_per_hour_pdf}
\end{subfigure}
\begin{subfigure}{0.24\textwidth}
  \centering
  \includegraphics[width=\linewidth]{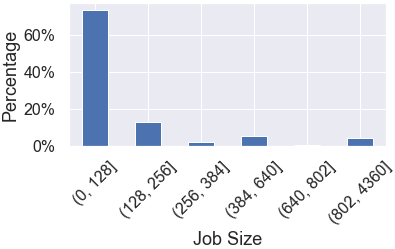}  
  \caption{Job size distribution.}
  \label{node_distribution_all}
\end{subfigure}
\begin{subfigure}{0.24\textwidth}
  \centering
  \includegraphics[width=\linewidth]{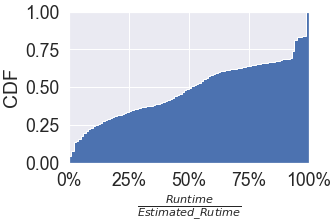}  
  \caption{Accuracy of job runtime estimates supplied by users.}
  \label{runtime_estimation_cdf}
\end{subfigure}
\caption{Job patterns of Theta training dataset.}
\label{analysis_training}
\end{figure}

We use 100 jobsets composed of 320,000 jobs for DRAS training on Theta. We collect 9 sampled jobsets by randomly selecting jobs from the original training trace and modeling job arrival times as Poisson distribution following the average inter-arrival time of the original trace. We split the original Theta training dataset into nine one-week jobsets. We generate 82 synthetic jobsets that mimic Theta workload patterns in terms of hourly and daily job arrivals, and distributions of job sizes and runtimes (Figure \ref{analysis_training}).

\begin{figure}[!h]
\centerline{\includegraphics[width=0.48\textwidth]{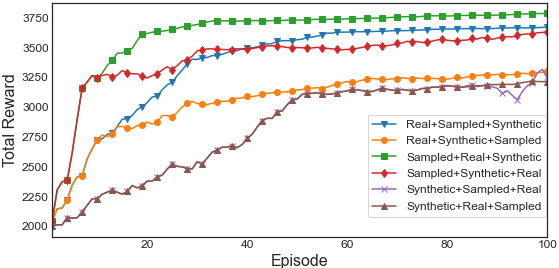}}
\caption{Comparison of quality and convergence of DRAS-PG by training it in different jobset orders (\S \ref{Training Strategy}). }
\label{order_total_reward}
\end{figure}

We validate the trained DRAS agent with an unseen validation dataset (i.e., March of 2018). Figure \ref{order_total_reward} compares the convergence rates by training DRAS in different jobset orderings. We make two key observations. First, training only with real jobsets (the first 9 episodes of the orange line) cannot obtain a converged model. To achieve a converged model, more jobsets are needed to train our agents. Second, training order plays an important role in performance. Training in the order of sampled, real and synthetic jobsets achieves the best result. While training with real jobsets first can also obtain a converged model, the performance is not as good as the case of training with sampled jobsets first. Training with synthetic jobsets first results in slow convergence. In summary, \textit{in order to generate a converged and high-quality model, DRAS needs to first learn from simple averaged cases (sampled jobsets) and then gradually move to more complicated special cases (real and synthetic jobsets).}

\begin{figure}[!h]
\centerline{\includegraphics[width=0.48\textwidth]{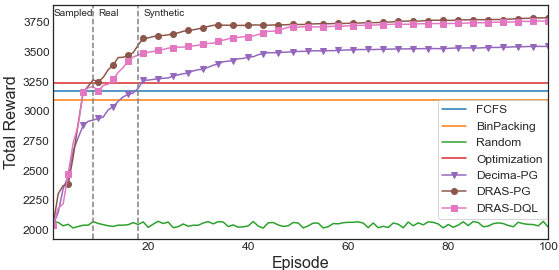}}
\caption{The total reward collected by the different scheduling methods on Theta validation dataset.}
\label{total_reward}
\end{figure}

Figure \ref{total_reward} shows the learning curves of different scheduling methods. Our three-phase training process allows DRAS to quickly learn and surpass other competing methods and converge to optimal solutions. Based on the results, we use the model trained after the 50th episode for testing listed in \S \ref{Evaluation}.

We perform a similar training and validation process on Cori. We train DRAS using 100 jobsets (20,000,000 jobs) composed of sampled traces, real traces, and synthetic traces. Both \textit{DRAS} methods converge at 40 episodes. Hence, we use the model trained after the 40th episode for testing.

\subsection{Evaluation Metrics}\label{evaluation metrics}
There are two classes of metrics for evaluating cluster scheduling: user-level metrics and system-level metrics. In our experiments, we measure four well-established metrics:
\begin{itemize}[leftmargin=*]
\item \textbf{Job wait time} is a user-level metric. It measures the interval between job submission to job start time. In our experiments, we analyze average job wait time, maximum job wait time, as well as the distribution of job wait times.
\item \textbf{Job response time} is a user-level metric which measures the interval between job submission to completion.
\item \textbf{Job slowdown} is another user-level metric. It measures the ratio of the job response time to its actual runtime. 
\item \textbf{System utilization} is a system-level metric. It measures the ratio of the used node-hours for useful job execution to the total elapsed node-hours.
\end{itemize}

\section{Experimental Results}\label{Evaluation}
Now we present the experimental results of applying the trained DRAS on test data (i.e., 21-month Theta log and 15-week Cori log). 
Our experiments intend to answer the following questions:
\begin{enumerate}
\item Does DRAS outperform existing scheduling? (\S \ref{Job Wait Time})
\item Is DRAS capable of preventing jobs from starvation? (\S \ref{Job Starvation Analysis})
\item If DRAS outperforms other methods, where does the performance gain come from? (\S \ref{Source of Performance Gains})
\item Can DRAS adapt to workload changes? (\S \ref{DRAS-DYN})
\end{enumerate}

\begin{figure}[!h]
\begin{subfigure}{0.5\textwidth}
  \centering
  \includegraphics[width=\linewidth]{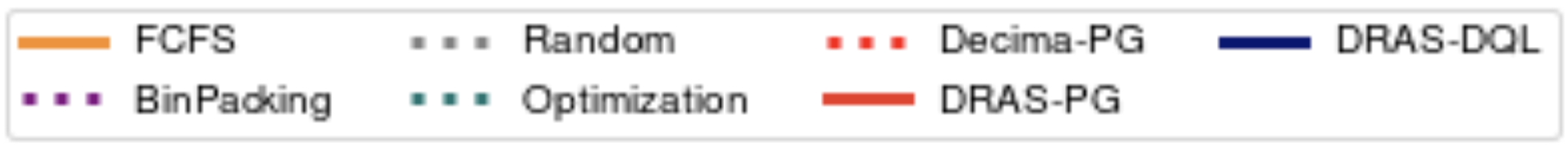}  
\end{subfigure}
\begin{subfigure}{0.24\textwidth}
  \centering
  \includegraphics[width=\linewidth]{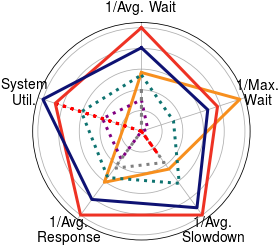}  
  \caption{Theta}
\end{subfigure}
\begin{subfigure}{0.24\textwidth}
  \centering
  \includegraphics[width=\linewidth]{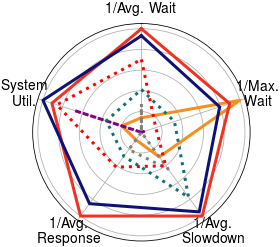}  
  \caption{Cori}
\end{subfigure}
\caption{Overall scheduling performance comparison using Kiviat graphs: Theta traces (left) and Cori traces (right). We use the reciprocal of average job wait time,  the reciprocal of maximum job wait time, the reciprocal of average slowdown, and the reciprocal of average job response time in the plots. All metrics are normalized to the range of 0 to 1. 1 means a method achieves the best performance among all methods and 0 means a method obtains the worst performance. The larger the area is, the better the overall performance is.}
\label{wait_times}
\end{figure}

\subsection{Scheduling Performance}\label{Job Wait Time} 
The quality of a scheduling method needs to be evaluated by multiple metrics, including both system-level and user-level metrics. Figure \ref{wait_times} presents the overall scheduling performance obtained by different scheduling methods. \textit{DRAS} yields the best result. \textit{DRAS-PG} achieves slightly better result on user-level metrics, while \textit{DRAS-DQL} obtains the best system-level performance. Although \textit{FCFS} has the lowest maximum wait time, it has poor performance on the rest of the metrics.  Both \textit{DRAS} methods outperform \textit{Optimization}, suggesting that \textit{DRAS} agents learn to select jobs that not only maximize the immediate reward, but also potentially improve performance in the future through maximizing cumulative reward. \textit{Decima-PG} achieves good performance on system utilization, but it fails to improve user-level metrics. \textit{BinPacking} and \textit{Random} have the worst performance, because they greedily select jobs one by one which ignores the best job combinations. Recall that \textit{DRAS} applies the similar strategy as \textit{Random} at the beginning of the training, by randomly exploring various actions, the better performance of \textit{DRAS} indicates that our RL models developed good policies through learning.

We also notice that some methods have inconsistent performance on the user-level metrics. For example, FCFS achieves the lowest maximum job wait time, however, it suffers from the high average job wait time. More detailed analysis of job wait time is needed. Due to the space limitation, we only present the in-depth analysis of job wait time on Theta in the following subsections.

\begin{figure}[!h]
\begin{subfigure}{0.5\textwidth}
  \centering
  \includegraphics[width=0.85\linewidth]{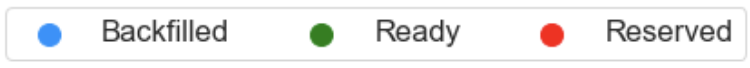}  
\end{subfigure}\\
\begin{subfigure}{0.24\textwidth}
  \centering
  \includegraphics[width=\linewidth]{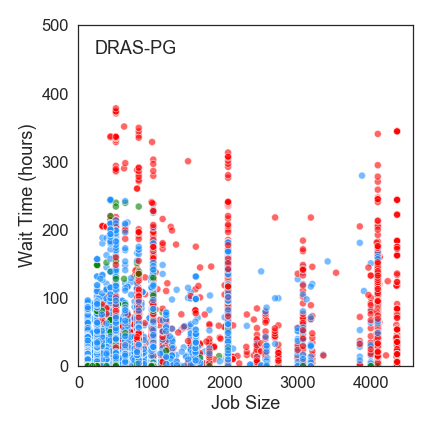}  
\end{subfigure}
\begin{subfigure}{0.24\textwidth}
  \centering
  \includegraphics[width=\linewidth]{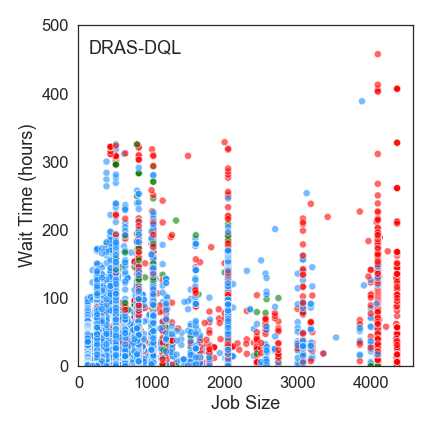}  
\end{subfigure}
\begin{subfigure}{0.24\textwidth}
  \centering
  \includegraphics[width=\linewidth]{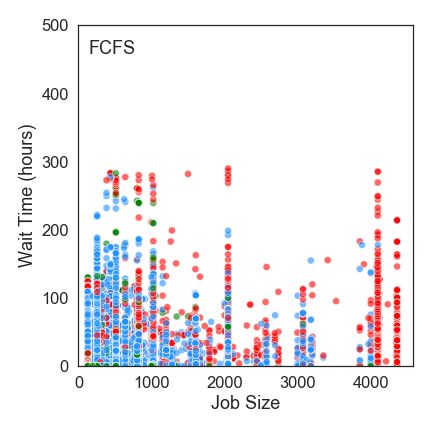}  
\end{subfigure}
\begin{subfigure}{0.24\textwidth}
  \centering
  \includegraphics[width=\linewidth]{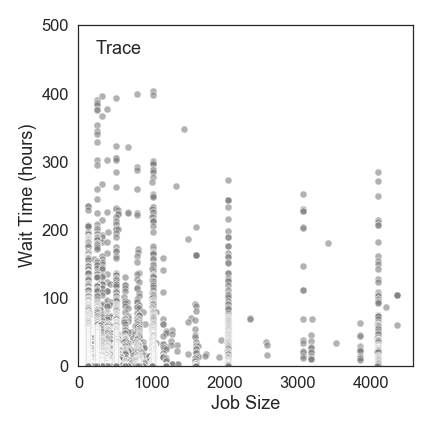}  
\end{subfigure}
\begin{subfigure}{0.24\textwidth}
  \centering
  \includegraphics[width=\linewidth]{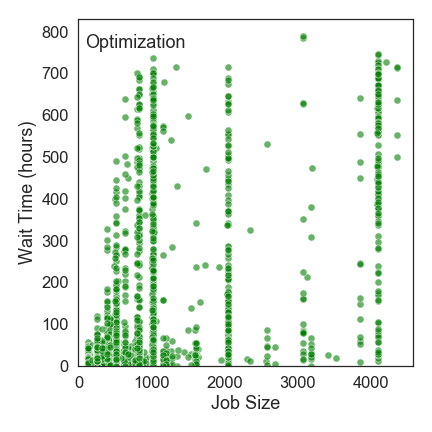}
\end{subfigure}
\begin{subfigure}{0.24\textwidth}
  \centering
  \includegraphics[width=\linewidth]{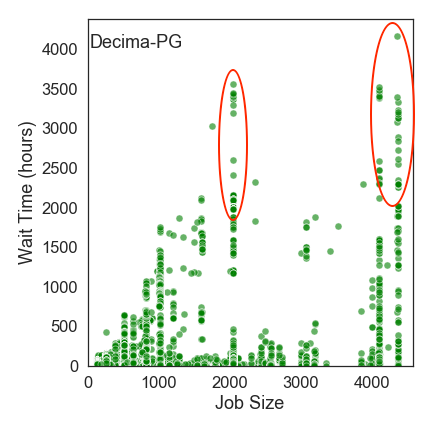}
\end{subfigure}
\begin{subfigure}{0.24\textwidth}
  \centering
  \includegraphics[width=\linewidth]{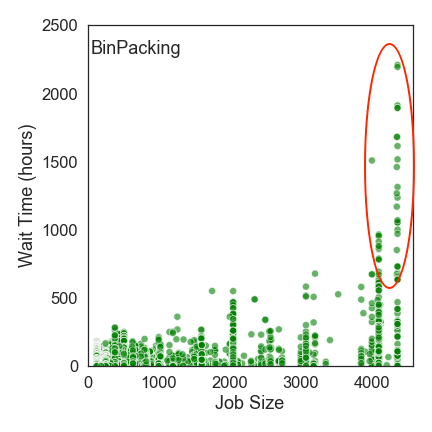}
\end{subfigure}
\begin{subfigure}{0.24\textwidth}
  \centering
  \includegraphics[width=\linewidth]{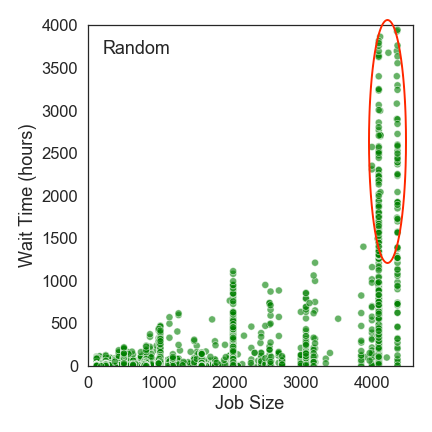}
\end{subfigure}
\caption{Job wait time distributions with respect to job size and job type on Theta. 
Note that the y-axis scale for \textit{Decima-PG}, \textit{BinPacking} and \textit{Random} is much larger than those for others. \textit{Trace} presents the job wait times extracted from the original log, which can be used as the baseline. Since we do not have the job type information, all jobs are marked in grey. Eclipses in the plots indicate \textit{Decima-PG}, \textit{BinPacking}, and \textit{Random} lead to severe job starvation. }
\label{job_size_vs_wait_time}
\end{figure}

\subsection{Job Starvation Analysis}\label{Job Starvation Analysis} 
Figure \ref{job_size_vs_wait_time} shows job wait times under different job sizes and categories. We make three key observations from this figure. First, \textit{DRAS} and \textit{FCFS} prevent jobs from starvation, while \textit{Decima-PG}, \textit{BinPacking} and \textit{Random} suffer severe job starvation. The maximum job wait times of \textit{DRAS-PG} and \textit{DRAS-DQL} are 16 days and 20 days respectively, which are only slightly higher than the maximum job wait time of \textit{FCFS} (13 days) and are similar to \textit{Trace}. Although \textit{Optimization} and \textit{DRAS} aim at the same scheduling objectives, the maximum wait time of \textit{Optimization} is twice as long as that of \textit{DRAS}. The maximum job wait times of \textit{Decima-PG}, \textit{BinPacking} and \textit{Random} are 170 days, 95 days, and 170 days, indicating they are not suitable for HPC cluster scheduling. Second, in \textit{Decima-PG}, \textit{BinPacking}, and \textit{Random}, large-sized jobs wait noticeably more time than small-sized jobs. They inherently give higher priority to small-sized jobs at the expense of large-sized jobs, because they lack reservation strategy to reserve resources for large-sized jobs. The bias toward small-sized jobs is not ideal for HPC scheduling, especially for capability systems. In contrast, the methods with reservation strategy, i.e., \textit{FCFS} and \textit{DRAS}, do not have a significant difference between small jobs' wait times and large jobs' wait times. This demonstrates that \textit{DRAS} and \textit{FCFS} are relatively fair scheduling policies. Third, if we take a look at the methods using reservation and backfilling strategies (i.e., \textit{FCFS} and \textit{DRAS}), we notice that almost all large jobs are executed through reservation, while the majority of small jobs are executed through backfilling. In short, these results demonstrate that DRAS is capable of preventing job starvation mainly due to the incorporation of job reservation and backfilling in our DRAS design.

\subsection{Source of DRAS Performance Gain}\label{Source of Performance Gains}
Table \ref{mode_distribution} presents job distributions by using different scheduling methods. We notice that although \textit{DRAS} backfills a majority of the jobs, most node hours are consumed by reserved jobs. If we read Table \ref{mode_distribution} along with Figure \ref{job_size_vs_wait_time}, we observe that there are a few jobs with wait time of over 300 hours and these jobs are mainly allocated through reservation by DRAS. Without reservation, these jobs would wait for 2X-10X more time as happened in \textit{Decima-PG}, \textit{Optimization}, \textit{BinPacking}, and \textit{Random}. 
Put together, these results reveal that DRAS learns to achieve the scheduling goals by prioritizing jobs and preventing job starvation through its reservation mechanism embedded in its two-level neural network design.

\begin{table}[!h]
\caption{Job distributions in different execution models (defined in \S \ref{Starvation and Backfill Management}) on Theta.}
\label{mode_distribution}
\resizebox{\linewidth}{!}{
\begin{tabular}{|l||l|l||l|l||l|l|}
\hline \rowcolor{Gray}
\multirow{2}{*}{} & \multicolumn{2}{l|}{Backfilled} & \multicolumn{2}{l|}{Ready} & \multicolumn{2}{l|}{Reserved} \\ \cline{2-7} 
\rowcolor{Gray}                  & jobs       & core hours    & jobs   & core hours & jobs      & core hours   \\ \hline\hline
\rowcolor{LightCyan} Optimization      & 0\%           & 0\%             & 100\%      & 100\%         & 0\%          & 0\%            \\ \hline
\rowcolor{Gray} Decima-PG         & 0\%           & 0\%             & 100\%      & 100\%         & 0\%          & 0\%            \\ \hline
\rowcolor{LightCyan} BinPacking        & 0\%           & 0\%             & 100\%      & 100\%         & 0\%          & 0\%            \\ \hline
\rowcolor{Gray} Random            & 0\%           & 0\%             & 100\%      & 100\%         & 0\%          & 0\%            \\ \hline
\rowcolor{LightCyan}  FCFS              & 79.25\%       & 30.45\%         & 9.88\%    & 16.99\%       & 10.87\%      & 52.56\%        \\ \hline
\rowcolor{Gray} DRAS-PG           & 83.76\%       & 33.67\%         & 8.63\%     & 11.29\%       & 7.61\%       & 55.04\%        \\ \hline
\rowcolor{LightCyan} DRAS-DQL          & 84.83\%       & 34.17\%         & 6.84\%     & 10.91\%       & 15.17\%       & 54.92\%        \\ \hline
\end{tabular}
}
\end{table}

\begin{figure}[!h]
\centerline{\includegraphics[width=0.5\textwidth]{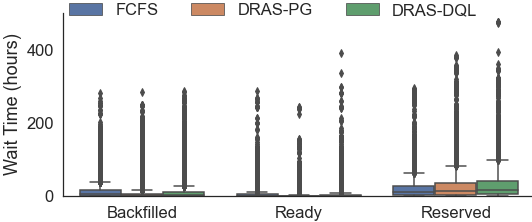}}
\caption{Bar plot of job wait times, grouping by job execution modes. As compared to \textit{FCFS}, \textit{DRAS} learns to intelligently select jobs for immediate execution, reservation, or backfilling so as to maximize the overall scheduling performance.}
\label{barplot_backfill}
\end{figure}

Although both \textit{FCFS} and \textit{DRAS} apply backfilling strategies, \textit{DRAS} performs significantly better in terms of average job wait time (Figure \ref{wait_times}). In Figure \ref{barplot_backfill}, we notice that DRAS largely reduces the wait time of ready and backfilled jobs at the expense of a slightly higher wait time for reserved jobs. \textit{FCFS} schedules jobs in their arrival order, while \textit{DRAS} selects jobs from the queue that aim to balance three objectives (i.e., minimizing average job wait time, prioritizing large jobs, and maximizing system utilization). Therefore, \textit{DRAS} learns to pick backfilled and ready jobs that lead to lower average job wait time and select jobs queued for long times to avoid job starvation. The better performance of \textit{DRAS} demonstrates that DRAS learns to intelligently select jobs for resource allocation so as to maximize the long-term scheduling performance.


\begin{figure}[!h]
  \centering
  \includegraphics[width=\linewidth]{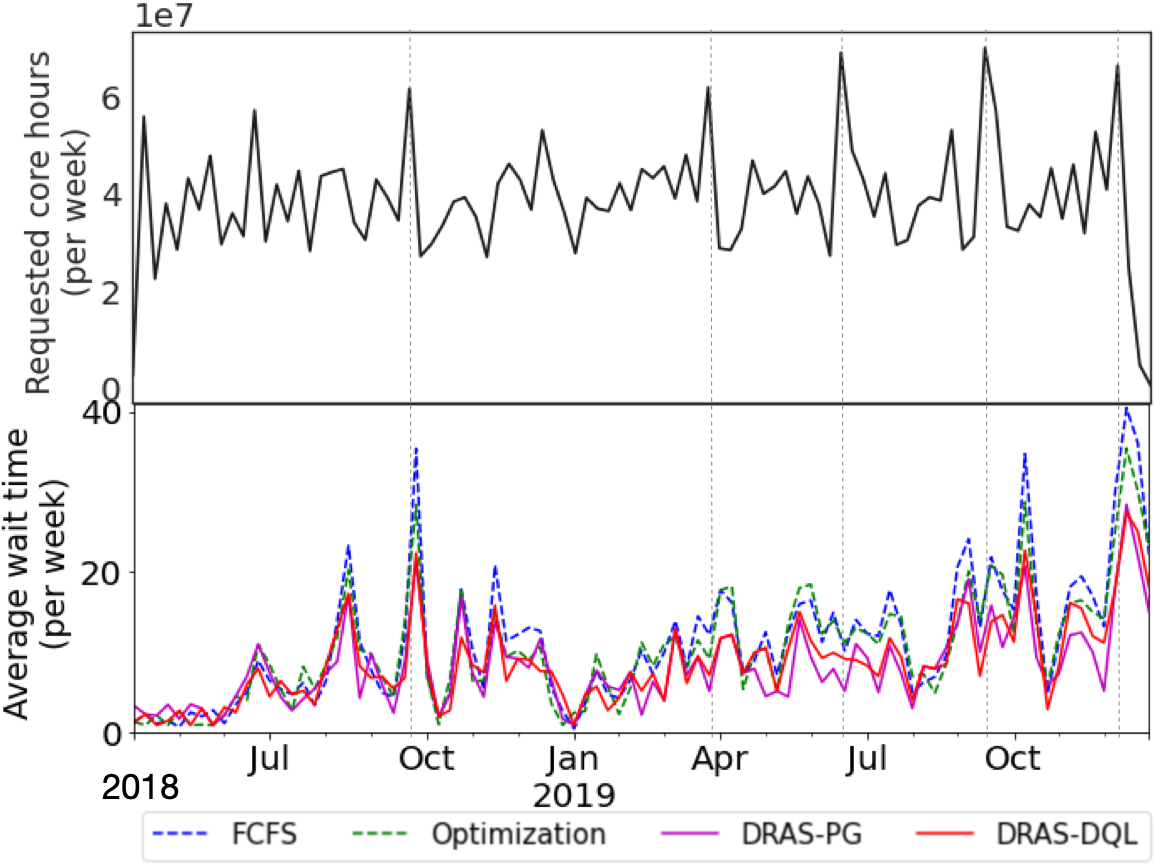}  
\caption{When the system load is high, DRAS dynamically adjusts it network parameters to reduce average job wait time. }
\label{dynamic}
\end{figure}

\subsection{Adaptation to Workload Change}\label{DRAS-DYN}

Figure \ref{dynamic} shows the total core hours and average job wait times per week during the testing period. The system loads are dynamically changing. Several dramatic demand surges put severe pressure on scheduling performance. The bottom figure compares how DRAS agents respond to the workload changes compared to other methods. It is clear that DRAS achieves greater wait time reduction when workload surges. Recall that DRAS agents continuously adjust their network parameters to minimize average job wait time. On the other hand, the policy of the static methods is predefined and fixed, which performs poorly under heavy workloads.

A system change, such as adding more nodes, requires DRAS to re-train the model. In our experiments, we spent less than 3 hours on a personal computer to obtain a converged model. Considering that system changes is not very frequent and DRAS can avoid complicated manually tuning policies, it is worth to re-train the model when the system changes.

\subsection{Runtime Overhead}
In our experiments, \textit{DRAS-PG} takes less than 1 second and \textit{DRAS-DQL} takes less than 2 seconds for each network parameter update during testing. Note that the experiments are conducted on a personal computer configured with Intel quad-core 2.6Ghz CPU with 16GB memory. In practice, HPC cluster scheduling is typically required to make decisions in 15-30 seconds \cite{Fan3}. In other words, the \textit{DRAS} agents impose trivial overhead, hence being feasible for online deployment.

\section{Conclusion}\label{Conclusion}
In this paper, we have presented DRAS, a RL-empowered HPC cluster scheduling agent. DRAS represents the scheduling policy as a hierarchical neural network and automatically learns customized policies through training with the system specific workloads.
Our results demonstrate that DRAS is capable of grasping system- and workload-specific characteristics, preventing
large jobs from starvation, adapting to workload changes without human intervention, and outperforming existing scheduling policies by up to 45\%. 
We hope it opens up exciting opportunities to rethink HPC cluster scheduling.

\section*{Acknowledgment}
This work is supported in part by US National Science Foundation grants CNS-1717763, CCF-1618776. 
T. Childers, W. Allcock, P. Rich, and M. Papka are supported by the Argonne Leadership Computing Facility, which is a U.S. Department of Energy Office of Science User Facility operated under contract DE-AC02-06CH11357. Job logs from the Cori system were provided by the National Energy Research Scientific Computing Center (NERSC), a U.S. Department of Energy Office of Science User Facility operated under Contract No. DE-AC02-05CH11231.

\balance
\bibliographystyle{unsrt}
\bibliography{ipdps1.bib}

\begin{thebibliography}{10}

\bibitem{Feitelson02}
A.~Mu'alem and D.~Feitelson.
\newblock {Utilization, Predictability, Workloads, and User Runtime Estimates
  in Scheduling the IBM SP2 with Backfilling}.
\newblock {\em TPDS'01}.

\bibitem{xu01}
X.~Yang, Z.~Zhou, S.~Wallace, Z.~Lan, W.~Tang, S.~Coghlan, and M.~Papka.
\newblock {Integrating Dynamic Pricing of Electricity into Energy Aware
  Scheduling for HPC Systems}.
\newblock SC'13.

\bibitem{Fan3}
Y.~Fan, Z.~Lan, P.~Rich, W.~Allcock, M.~Papka, B.~Austin, and D.~Paul.
\newblock {Scheduling Beyond CPUs for HPC}.
\newblock HPDC'19.

\bibitem{Sun}
H.~Sun, P.~Stolf, J.~Pierson, and G.~Costa.
\newblock {Energy-Efficient and Thermal-Aware Resource Management for
  Heterogeneous Datacenters}.
\newblock In {\em Sustainable Computing: Informatics and Systems}, 2014.

\bibitem{Qiao1}
P.~Qiao, X.~Wang, X.~Yang, Y.~Fan, and Z.~Lan.
\newblock {Preliminary Interference Study About Job Placement and Routing
  Algorithms in the Fat-Tree Topology for HPC Applications}.
\newblock In {\em CLUSTER}, 2017.

\bibitem{Qiao2}
P.~Qiao, X.~Wang, X.~Yang, Y.~Fan, and Z.~Lan.
\newblock {Joint Effects of Application Communication Pattern, Job Placement
  and Network Routing on Fat-Tree Systems}.
\newblock In {\em ICPP Workshops}, 2018.

\bibitem{Grandl2014}
R.~Grandl, G.~Ananthanarayanan, S.~Kandula, S.~Rao, and A.~Akella.
\newblock {Multi-resource Packing for Cluster Schedulers}.
\newblock SIGCOMM'14.

\bibitem{Mao2019}
H.~Mao, M.~Schwarzkopf, S.~Venkatakrishnan, Z.~Meng, and M.~Alizadeh.
\newblock {Learning Scheduling Algorithms for Data Processing Clusters}.
\newblock SIGCOMM'19.

\bibitem{Ahmad2017}
A.~Sallab, M.~Abdou, E.~Perot, and S.~Yogamani.
\newblock {Deep Reinforcement Learning framework for Autonomous Driving}.
\newblock {\em Electronic Imaging}.

\bibitem{Johannink2019}
T.~{Johannink}, S.~{Bahl}, A.~{Nair}, J.~{Luo}, A.~{Kumar}, M.~{Loskyll},
  J.~{Ojea}, E.~{Solowjow}, and S.~{Levine}.
\newblock {Residual Reinforcement Learning for Robot Control}.
\newblock ICRA'19.

\bibitem{Mnih2013PlayingAW}
V.~Mnih, K.~Kavukcuoglu, D.~Silver, A.~Graves, I.~Antonoglou, D.~Wierstra, and
  M.~Riedmiller.
\newblock {Playing Atari with Deep Reinforcement Learning}.
\newblock {\em NIPS Deep Learning Workshop}, 2013.

\bibitem{Go2017}
D.~Silver, J.~Schrittwieser, K.~Simonyan, I.~Antonoglou, A.~Huang, A.~Guez,
  T.~Hubert, L.~Baker, M.~Lai, A.~Bolton, Y.~Chen, T.~Lillicrap, F.~Hui,
  L.~Sifre, G.~Driessche, T.~Graepel, and D.~Hassabis.
\newblock {Mastering the Game of Go without Human Knowledge}.
\newblock {\em Nature}, 2017.

\bibitem{RL2017}
R.~Sutton and A.~Barto.
\newblock {Reinforcement Learning: An Introduction, Second Edition}.
\newblock {\em MIT Press}, 2017.

\bibitem{Fan2}
W.~Allcock, P.~Rich, Y.~Fan, and Z.~Lan.
\newblock {Experience and Practice of Batch Scheduling on Leadership
  Supercomputers at Argonne}.
\newblock JSSPP'17.

\bibitem{Topper}
L.~Yu, Z.~Zhou, Y.~Fan, M.~Papka, and Z.~Lan.
\newblock {System-wide Trade-off Modeling of Performance, Power, and Resilience
  on Petascale Systems}.
\newblock In {\em The Journal of Supercomputing}, 2018.

\bibitem{SLURM}
M.~Jette, A.~Yoo, and M.~Grondona.
\newblock {SLURM: Simple Linux Utility for Resource Management}.
\newblock In {\em JSSPP'03}.

\bibitem{Moab}
{Moab}.
\newblock http://www.adaptivecomputing.com/
  products/hpc-products/moab-hpc-basic-edition/.

\bibitem{PBS}
{PBS Professional}.
\newblock http://www.pbsworks.com/.

\bibitem{Cobalt}
{Cobalt}.
\newblock https://www.alcf.anl.gov/cobalt-scheduler.

\bibitem{Fan1}
Y.~Fan, P.~Rich, W.~Allcock, M.~Papka, and Z.~Lan.
\newblock {Trade-Off Between Prediction Accuracy and Underestimation Rate in
  Job Runtime Estimates}.
\newblock CLUSTER'17.

\bibitem{Fan4}
Y.~Fan and Z.~Lan.
\newblock {Exploiting Multi-Resource Scheduling for HPC}.
\newblock In {\em SC Poster}, 2019.

\bibitem{Mao2016}
H.~Mao, M.~Alizadeh, I.~Menache, and S.~Kandula.
\newblock {Resource Management with Deep Reinforcement Learning}.
\newblock HotNets'16.

\bibitem{RLScheduler}
D.~Zhang, D.~Dai, Y.~He, F.~Bao, and B.~Xie.
\newblock {RLScheduler: An Automated HPC Batch Job Scheduler Using
  Reinforcement Learning}.
\newblock SC'20.

\bibitem{Ipek2008}
E.~{\.Ipek}, O.~{Mutlu}, J.~{Mart\'inez}, and R.~{Caruana}.
\newblock {Self-Optimizing Memory Controllers: A Reinforcement Learning
  Approach}.
\newblock ISCA'08.

\bibitem{DL2015}
Y.~LeCun, Y.~Bengio, and G.~Hinton.
\newblock {Deep Learning}.
\newblock {\em Nature}, 2015.

\bibitem{Sutton1999}
R.~Sutton, D.~McAllester, S.~Singh, and Y.~Mansour.
\newblock {Policy Gradient Methods for Reinforcement Learning with Function
  Approximation}.
\newblock NIPS'99.

\bibitem{Mnih2016}
V.~Mnih, A.~Badia, M.~Mirza, A.~Graves, T.~Harley, T.~Lillicrap, D.~Silver, and
  K.~Kavukcuoglu.
\newblock {Asynchronous Methods for Deep Reinforcement Learning}.
\newblock ICML'16.

\bibitem{CoriPolicy}
{NERSC Queue Policies}.
\newblock https://docs.nersc.gov/jobs/policy/.

\bibitem{Goodfellow-et-al-2016}
I.~Goodfellow, Y.~Bengio, and A.~Courville.
\newblock {\em {Deep Learning}}.
\newblock MIT Press.

\bibitem{Adam}
D.~Kingma and J.~Ba.
\newblock {Adam: A Method for Stochastic Optimization}.
\newblock ICLR'15.

\bibitem{TensorflowURL}
{Tensorflow}.
\newblock https://www.tensorflow.org/.

\bibitem{DRASGithub}
{DRAS Github Repository}.
\newblock https://github.com/SPEAR-IIT/CQSim/tree/DRAS.

\bibitem{Li1}
B.~Li, S.~Chunduri, K.~Harms, Y.~Fan, and Z.~Lan.
\newblock {The Effect of System Utilization on Application Performance
  Variability}.
\newblock In {\em ROSS}, 2019.

\bibitem{CQSimGithub}
{CQSim Github Repository}.
\newblock https://github.com/SPEAR-IIT/CQSim.

\bibitem{Theta}
{Theta}.
\newblock https://www.alcf.anl.gov/theta.

\bibitem{ThetaPolicy}
{Job Scheduling Policy for Theta}.
\newblock
  https://www.alcf.anl.gov/support-center/theta/job-scheduling-policy-theta.

\bibitem{Cori}
{Cori}.
\newblock https://docs.nersc.gov/systems/cori/.

\end{thebibliography}

\end{document}